\documentclass[12pt]{article}
\usepackage{cite,epsfig}

\input paperdef

\begin{document}

\thispagestyle{empty}
\setcounter{page}{0}
\def\thefootnote{\fnsymbol{footnote}}

\begin{flushright}
CERN-TH/2000-55\\
DESY 00-020\\
KA-TP-3-2000\\
hep-ph/0002213 \\
\end{flushright}

\vspace{1cm}

\begin{center}

{\large\sc {\bf \fhf: a program for a fast calculation}}

\vspace*{0.4cm} 

{\large\sc {\bf of masses and mixing angles}}

\vspace*{0.4cm}

{\large\sc {\bf in the Higgs Sector of the MSSM}}
\footnote{Extended version of the contribution to the 
          {\em Report of the HIGGS working group for the
                 Workshop ``Physics at TeV Colliders''}, 
                 Les Houches, France, June 1999.}
 
\vspace{1cm}

{\sc 
S.~Heinemeyer$^{1}$%
\footnote{email: Sven.Heinemeyer@desy.de}%
, W.~Hollik$^{2}$%
\footnote{email: Wolfgang.Hollik@physik.uni-karlsruhe.de}%
~and G.~Weiglein$^{3}$%
\footnote{email: Georg.Weiglein@cern.ch}
}

\vspace*{1cm}

{\sl
$^1$ 
DESY Theorie, Notkestr. 85, 22603 Hamburg, Germany

\vspace*{0.4cm}

$^2$ Institut f\"ur Theoretische Physik, Universit\"at Karlsruhe, \\
D--76128 Karlsruhe, Germany

\vspace*{0.4cm}

$^3$ CERN, TH Division, CH-1211 Geneva 23, Switzerland
}

\end{center}

\vspace*{1cm}

\begin{abstract}
\fhf\ is a Fortran code for the calculation of the masses and the
mixing angle of the
neutral $\cp$-even Higgs bosons in the MSSM up to \twol\ order. It is based
on a compact analytical approximation formula of the complete
diagrammatic \onel\ and the dominant \twol\ contributions.
At the \onel\ level a
leading logarithmic result is used, taking into account all sectors of
the MSSM. At the \twol\ level at $\oaas$ the leading logarithmic and
non-logarithmic contributions are taken into account. 
The approximation formula is valid for 
arbitrary choices of the parameters in the Higgs sector of the model.
Comparing its quality to the full diagrammatic
result, we find agreement better than $2 \gev$ for
most parts of the MSSM parameter space.
\end{abstract}

\def\thefootnote{\arabic{footnote}}
\setcounter{page}{0}
\setcounter{footnote}{0}

\newpage


\section{Introduction}

One of the most striking phenomenological implications of Supersymmetry
(SUSY) is the prediction of a relatively light Higgs boson, which is 
common to all supersymmetric models whose couplings remain in the
perturbative regime up to a very high energy
scale~\cite{susylighthiggs}. The search for the lightest Higgs boson thus 
allows a crucial test of SUSY and especially the Minimal
Supersymmetric Standard Model (MSSM)~\cite{mssm}. Therefore it is one
of the main 
goals at the present and the next generation of colliders. 
A precise knowledge of the dependence of the masses and mixing angles
of the Higgs sector of the MSSM
on the relevant SUSY parameters is necessary for a
detailed analysis of SUSY phenomenology at LEP2, the upgraded Tevatron,
and also for the LHC and a future linear $e^+e^-$ collider, where
high-precision measurements in the Higgs sector might become possible.

The mass of the lightest $\cp$-even MSSM Higgs boson, $\mh$, is
bounded from above by about $\mh \lsim 135 \gev$, including radiative
corrections up to the \twol\ level~%
\cite{mhiggsletter,mhiggslong,mhiggsRG1a,mhiggsRG1b,mhiggsRG2}.
The mixing angle $\aeff$ that diagonalizes the neutral $\cp$-even Higgs
sector receives the same kind of corrections. By incorporating
$\aeff$ into the Higgs decay widths, the leading electroweak corrections
to the decay of the neutral $\cp$-even Higgs bosons are taken into account.

In this note we present the Fortran code \fhf. Using low
energy MSSM parameters as input, it evaluates the masses of the neutral
$\cp$-even Higgs bosons, $\mh$ and $\mH$, as well as the corresponding
mixing angle, $\aeff$, at the \twol\ level. In addition the mass of
the charged Higgs boson, $\mhp$, is evaluated at the \onel\
level. The \rp, leading to constraints in the scalar fermion
sector of the MSSM, is evaluated up to $\oaas$, taking into account the
gluon exchange contribution at the \twol\ level~\cite{drhosuqcd}.
\fhf\ is based on a compact analytical approximation formula,
containing at the \twol\ level the dominant corrections in
$\oaas$ obtained in the Feynman-diagrammatic (FD)
approach~\cite{mhiggsletter,mhiggslong} and subdominant corrections of
$\ogmzmts$ obtained with renormalization group (RG)
methods~\cite{mhiggsRG1a,mhiggsRG1b,mhiggsRG2}.
In comparison with the FD result, consisting of the complete
\onel\ and the \twol\ contributions as given in \citere{mhiggslong}
(incorporated into the Fortran code \fh~\cite{feynhiggs}), the
approximation formula is much shorter. Thus \fhf\ is about $3 \times 10^4$
times faster than \fh. Agreement between these two codes of
better than $2 \gev$ is found for most parts of the MSSM parameter
space. 

The following sections are organized as follows: In Sect.~2 we 
shortly summarize the analytical approximation formula and perform a
comparison with the full FD result.
A description of how to use \fhf\ is given in
Sect.~3. The conclusions can be found in Sect.~4.


\section{The analytical approximation}

In order to fix our notations, 
we first list the conventions for the
MSSM scalar top sector:
the mass matrix in the basis of the current eigenstates $\StopL$ and
$\StopR$ is given by (neglecting contributions $\sim \MZ^2$):
\BE
\label{stopmassmatrixwithdt}
{\cal M}^2_{\Stop} =
  \ML \MstL^2 + \mt^2  &
      \mt \Xt \\
      \mt \Xt &
      \MstR^2 + \mt^2  
  \MR,
\EE
where 
\BE
\mt \Xt = \mt (A_t - \mu \CTb)~.
\label{eq:mtlr}
\EE
Furtheron the following simplification is used:
\BE
\MstL = \MstR := \msq .
\EE
In this simplified case we define%
\footnote{
The more general case with $\MstL \neq \MstR$ can be found in
\refeq{msfkt}. 
}
\BE
\ms^2 := \msq^2 + \mt^2 .
\EE
For further details see \citere{mhiggslle}.

At the tree level the mass matrix of the neutral $\cp$-even Higgs bosons
in the $\phi_1,\phi_2$ basis can be expressed as follows:
\BEA
M_{\rm Higgs}^{2, {\rm tree}} &=& \ML \mpe^2 & \mpez^2 \\ 
                           \mpez^2 & \mpz^2 \MR \non\\
&=& \ML \MA^2 \SQb + \MZ^2 \CQb & -(\MA^2 + \MZ^2) \Sb \Cb \\
    -(\MA^2 + \MZ^2) \Sb \Cb & \MA^2 \CQb + \MZ^2 \SQb \MR.
\label{higgsmassmatrixtree}
\EEA
$\MA$ denotes the mass of the $\cp$-odd Higgs boson, $\MZ$ is the mass
of the $Z$ boson, and $\tb = v_2/v_1$ is the ratio of the two vacuum
expectation values of the two Higgs doublets in the MSSM, see \citere{hhg}.

At higher orders, the Higgs mass matrix (\ref{higgsmassmatrixtree}) is
supplemented by the renormalized self-energies 
$\hSi_s(q^2)\,,\;s = \Pe,\Pz,\PePz$. Here we use the approximation of
neglected external 
momentum: $\hSi_s \equiv \hSi_s(0)$. The masses $\mh$ and
$\mH$ are then 
obtained by diagonalizing the higher order corrected mass matrix with
the mixing angle $\aeff$:
\BEA
&&
\label{higgsmassmatrix}
\ML \mpe^2 - \hSi_{\Pe} & \mpez^2 - \hSi_{\PePz} \\ 
    \mpez^2 - \hSi_{\PePz} & \mpz^2 - \hSi_{\Pz} \MR 
   \stackrel{\aeff}{\longrightarrow}
   \ML \mH^2 & 0 \\ 0 &  \mh^2 \MR~, \\
\label{alphaeff}
&& \aeff = {\rm arctan}\KKL 
  \frac{-(\MA^2 + \MZ^2) \Sb \Cb - \hSi_{\PePz}}
       {\MZ^2 \CQb + \MA^2 \SQb - \hSi_{\Pe} - \mh^2} \KKR~,~~
  -\frac{\pi}{2} < \aeff < \frac{\pi}{2}~.
\EEA



Here we only give a very brief outline of the calculation of the
renormalized Higgs boson self-energies. A detailed description can be
found in \citere{mhiggslle}.

The mass of the lightest Higgs boson receives contributions from all
sectors of the MSSM, but not all are numerically of equal relevance.
The dominant corrections arise from the $t-\Stop$ sector of the MSSM: 
The results for the renormalized self-energies $\hSi_s$ have been
derived analytically in the FD approach in
\citeres{mhiggsletter,mhiggslong}. These results, however, are rather
lengthy. In order to derive a compact analytical expression, 
several approximations have been made as described in \citere{mhiggslle}.
The main step of the approximation consists of a Taylor
expansion in 
\BE
\dst = \frac{|\mt\,\Xt|}{\ms^2}
\label{deltastop}
\EE
of the $\hSi_s(0)$.
For the \onel\ correction the expansion has been performed up to 
${\cal O}(\dst^8)$; 
all three renormalized \onel\ Higgs-boson self-energies give a contribution.
Concerning the \twol\ self-energies,
the expansion has been carried out
up to ${\cal O}(\dst^4)$; 
in the approximation considered here only $\hSiz_{\Pz}$ gives a
non-zero contribution. 

Leading contributions beyond $\oaas$ have been taken into account by
incorporating the leading \twol\ Yukawa 
correction of ${\cal O}(\gf^2\mt^6)$
\cite{mhiggsRG1a,mhiggsRG1b,mhiggsRG2}. Furthermore the result has
been expressed in terms of the \msbar\ top-quark mass
\BE
\mtms = \mtms(\mt) \approx \frac{\mt}{1 + \frac{4}{3\,\pi} \als(\mt)}~
\label{mtrun}
\EE
instead of the pole mass $\mt$.
This leads to an additional contribution in 
${\cal O}(\al\als^2)$.

The analytical expressions arising from the $t-\Stop$ sector are
given at the \onel\ level as follows:
\BEA
\label{p1seol}
\hSie_{\Pe}(0) &=&
    \frac{\gf\wz}{\pi^2} \MZ^4 \Lambda \CQb
    \log\lmtmsms~, \\
\label{p1p2seol}
\hSie_{\PePz}(0) &=&
    -\frac{\gf\wz}{\pi^2} \MZ^2 \CTb 
    \KKL -\frac{3}{8} \mtms^2 + \MZ^2 \Lambda \SQb \KKR \log\lmtmsms~, \\
\label{p2seol}
\hSie_{\Pz}(0) &=&
    \frac{\gf\wz}{\pi^2} \frac{\mtms^4}{8 \SQb} 
      \Bigg\{ -2 \frac{\MZ^2}{\mtms^2} 
              + \frac{11}{10}\frac{\MZ^4}{\mtms^4} \non \\
 && {} + \KKL 12 - 6 \frac{\MZ^2}{\mtms^2} \SQb +
                 8 \frac{\MZ^4}{\mtms^4} \Lambda \SVb \KKR \log\lmtmsms \non \\
 && {} + \frac{\Xt^2}{\ms^2}\, 
      \KKL -12 + 4 \frac{\MZ^2}{\mtms^2}
          + 6 \frac{\mtms^2}{\ms^2} \KKR 
    + \frac{\Xt^4}{\ms^4}\,
      \KKL 1 - 4\,\frac{\mtms^2}{\ms^2} 
           +3 \frac{\mtms^4}{\ms^4} \KKR \non \\
 && {} + \frac{\Xt^6}{\ms^6}\,
      \KKL  \frac{3}{5}\frac{\mtms^2}{\ms^2}
           -\frac{12}{5}\frac{\mtms^4}{\ms^4}
           +2\frac{\mtms^6}{\ms^6} \KKR \non \\
 && {} + \frac{\Xt^8}{\ms^8}\,
      \Bigg[  \frac{3}{7}\frac{\mtms^4}{\ms^4}
             -\frac{12}{7}\frac{\mtms^6}{\ms^6}
             +\frac{3}{2}\frac{\mtms^8}{\ms^8} \Bigg]
    \Bigg\}~, 
\EEA
with
\BE
\Lambda = \left(\frac{1}{8} - \frac{1}{3} \sw^2 + \frac{4}{9} 
                 \sw^4 \right), \qquad \sw^2 = 1 - \frac{\MW^2}{\MZ^2} . 
\label{eq:lambda}
\EE
The \twol\ contributions in $\oaas$ read:
\BEA
\hSiz_{\Pe}(0) &=& 0~, \non \\
\hSiz_{\PePz}(0) &=& 0~, \non \\
\label{p2setl}
\hSiz_{\Pz}(0) &=&
    \frac{\gf\wz}{\pi^2} \frac{\als}{\pi} \frac{\mtms^4}{\SQb}
      \Biggl[ 4 + 3 \log^2\lmtmsms + 2 \log\lmtmsms 
             -6 \frac{\Xt}{\ms} \non \\
 && {}  
     - \frac{\Xt^2}{\ms^2} \KKKL 3 \log\lmtmsms +8 \KKKR
     +\frac{17}{12} \frac{\Xt^4}{\ms^4} \Biggr]~.
\EEA
The \twol\ Yukawa correction in this approximation reads:
\BE
\label{mh2yuk}
\hSiz_{\Pz}(0) = -\frac{9}{16\pi^4} \frac{\gf^2 \mtms^6}{\SQb}
               \KKL \log^2\lmtmsms 
                    - 2 \frac{\Xt^2}{\ms^2} \log\lmtmsms
                    + \ed{6} \frac{\Xt^4}{\ms^2} \log\lmtmsms \KKR~.
\EE
$\ms$ has to be chosen according to
\BE
\label{msfkt}
\ms = \KKKL \begin{array}{l@{\quad:\quad}l}
            \sqrt{\msq^2 + \mtms^2} & \MstL = \MstR = \msq \\
            \KKL \MstL^2 \MstR^2 + \mtms^2 (\MstL^2 + \MstR^2) +
                     \mtms^4 \KKR^\frac{1}{4} &
                                    \MstL \neq \MstR
            \end{array} \right. 
\EE

For the \onel\ corrections from the other sectors of the MSSM 
the logarithmic approximation given in \citere{mhiggs1lrest} has been
used for \fhf, see also \citere{mhiggslle}. 
In order to obtain the radiatively corrected Higgs boson masses and
the mixing angle, the renormalized Higgs boson self-energies have to be
inserted into \refeq{higgsmassmatrix}, 
and the corresponding diagonalization has to be performed.


\bigskip

We have also implemented the result of the MSSM contributions to
$\De\rho$~\cite{drhosuqcd,precobssusyqcd}. Here the corrections 
arising from $\Stop/\Sbot$-loops up to $\oaas$
have been taken into account. The result is valid for arbitrary
parameters in  
the $\Stop$- and $\Sbot$-sector, also taking into account the mixing
in the $\Sbot$-sector which can have a non-negligible
effect in the large $\tb$ scenario~\cite{precobssusyqcd}.  

The \twol\ result can be separated into the pure gluon-exchange contribution,
which can be expressed by a very compact formula, allowing a very
fast evaluation, and the pure
gluino-exchange contribution, which is given by a rather lengthy 
expression. The latter correction goes to zero with increasing gluino mass
and can thus be discarded for a heavy gluino.
Concerning the implementation of the \rp\ into \fhf, we have neglected the
gluino exchange contribution.
The \rp\ can be used as an additional constraint (besides
the experimental bounds) on the squark masses. 
A value of $\De\rho$ outside the experimentally preferred region of 
$\dr^{\SU} \approx 10^{-3}$~\cite{delrhoexp} indicates experimentally
disfavored $\Stop$- and $\Sbot$-masses.



\bigskip
For illustration of the quality of the compact approximation,
we compare $\mh$ and $\aeff$ with the results from the complete
Feynman-diagrammatic calculation.
The FD calculation contains the full diagrammatic \onel\
contribution~\cite{mhiggs1lfull}, the complete leading \twol\
corrections in $\oaas$~\cite{mhiggsletter,mhiggslong},
and the two contributions beyond $\oaas$ (see \refeq{mtrun} 
and \refeq{mh2yuk}) without approximation.

\begin{figure}[ht!]
\begin{center}
 \mbox{
\psfig{figure=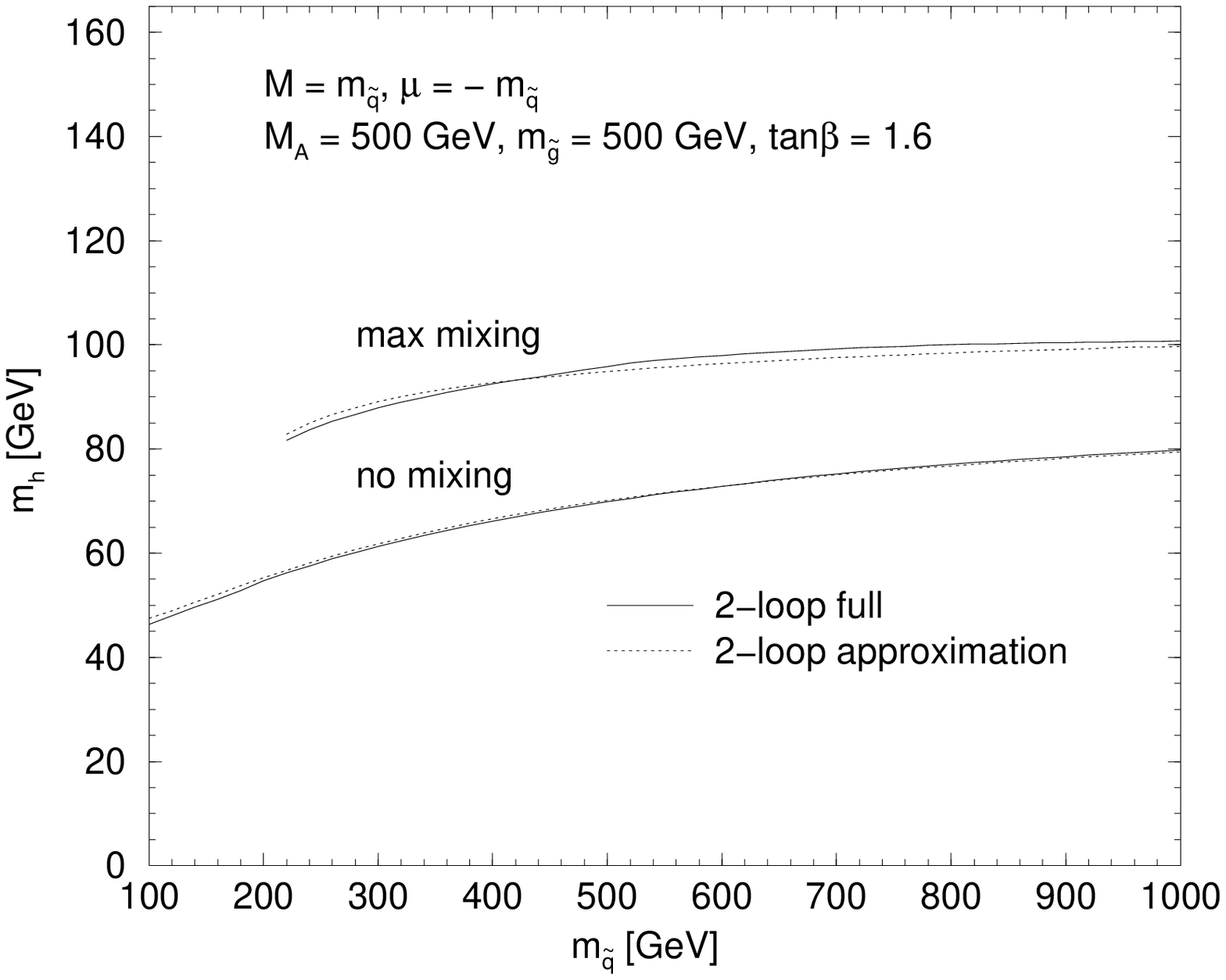,width=7cm,height=5.4cm}}
\hspace{1em}
\mbox{
\psfig{figure=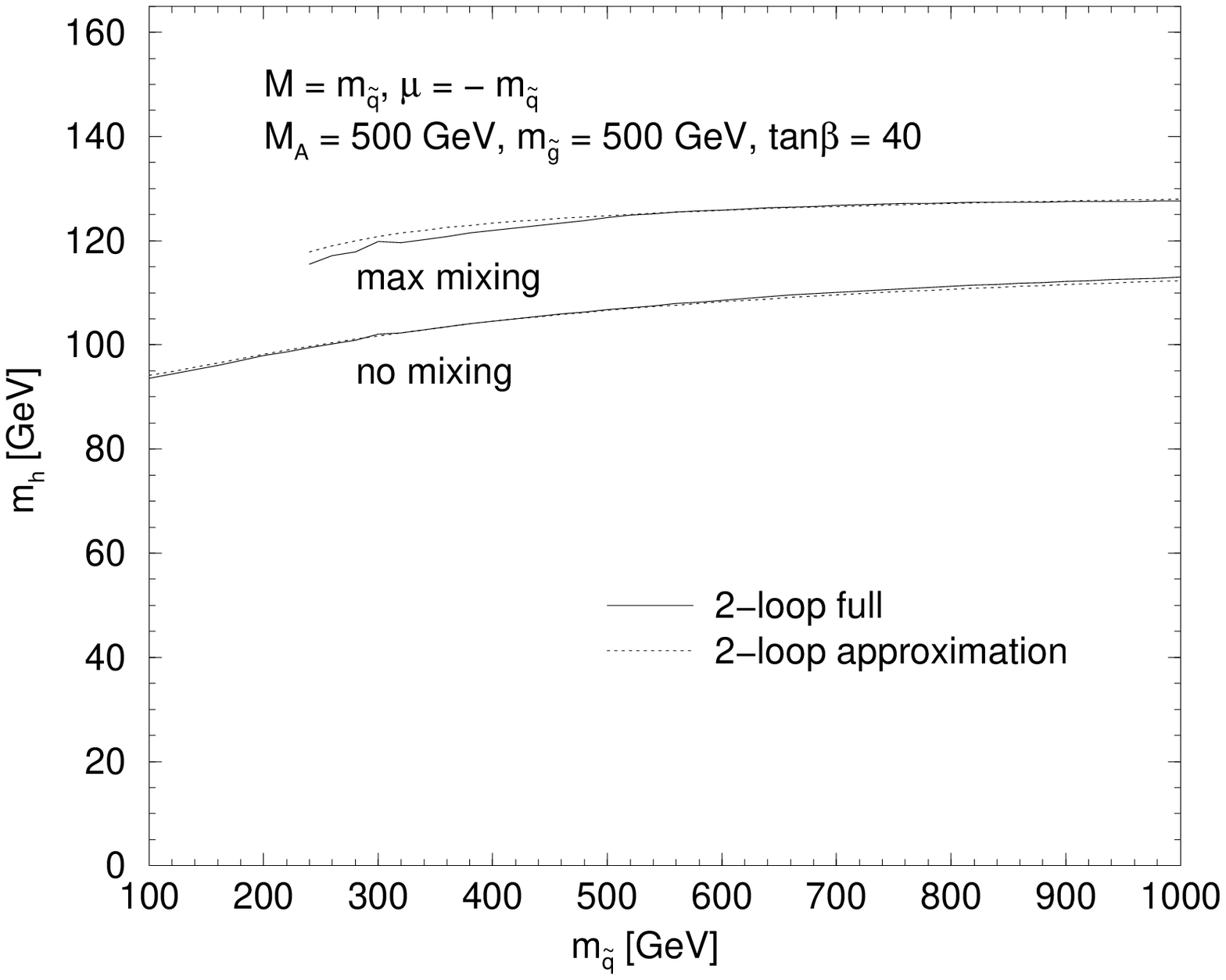,width=7cm,height=5.4cm}}
\end{center}
\vspace{-1.5em}
\caption[]{
$\mh$ as a function of $\msq$, calculated from the full formula and
from the approximation formula for $\MA = 500 \gev, 
\mgl = 500 \gev$ and $\tb = 1.6$ or $40$.
}
\label{fig:mh_mq}
\end{figure}
%
\begin{figure}[ht!]
\begin{center}
\mbox{
\psfig{figure=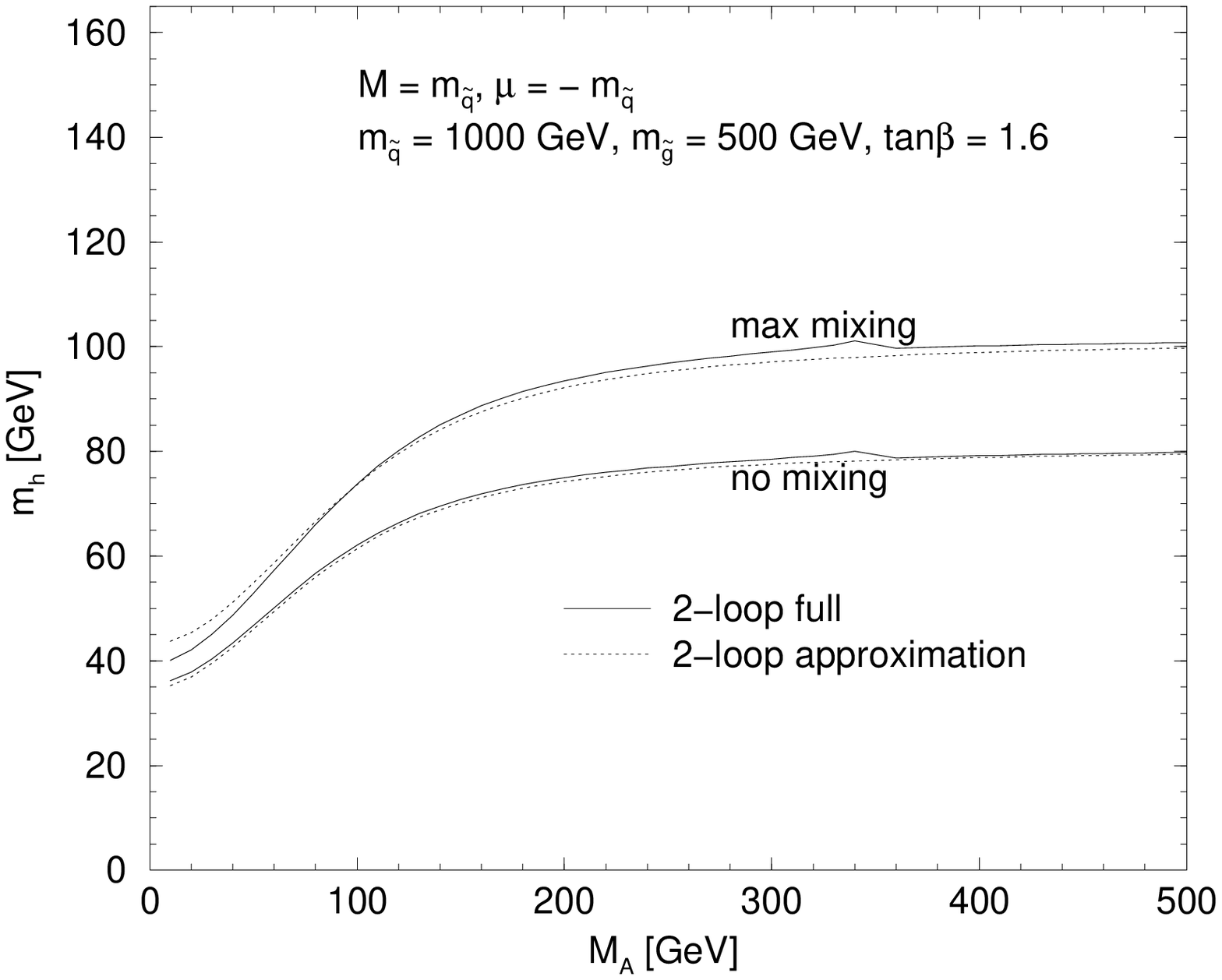,width=7cm,height=5.4cm}}
\hspace{1em}
\mbox{
\psfig{figure=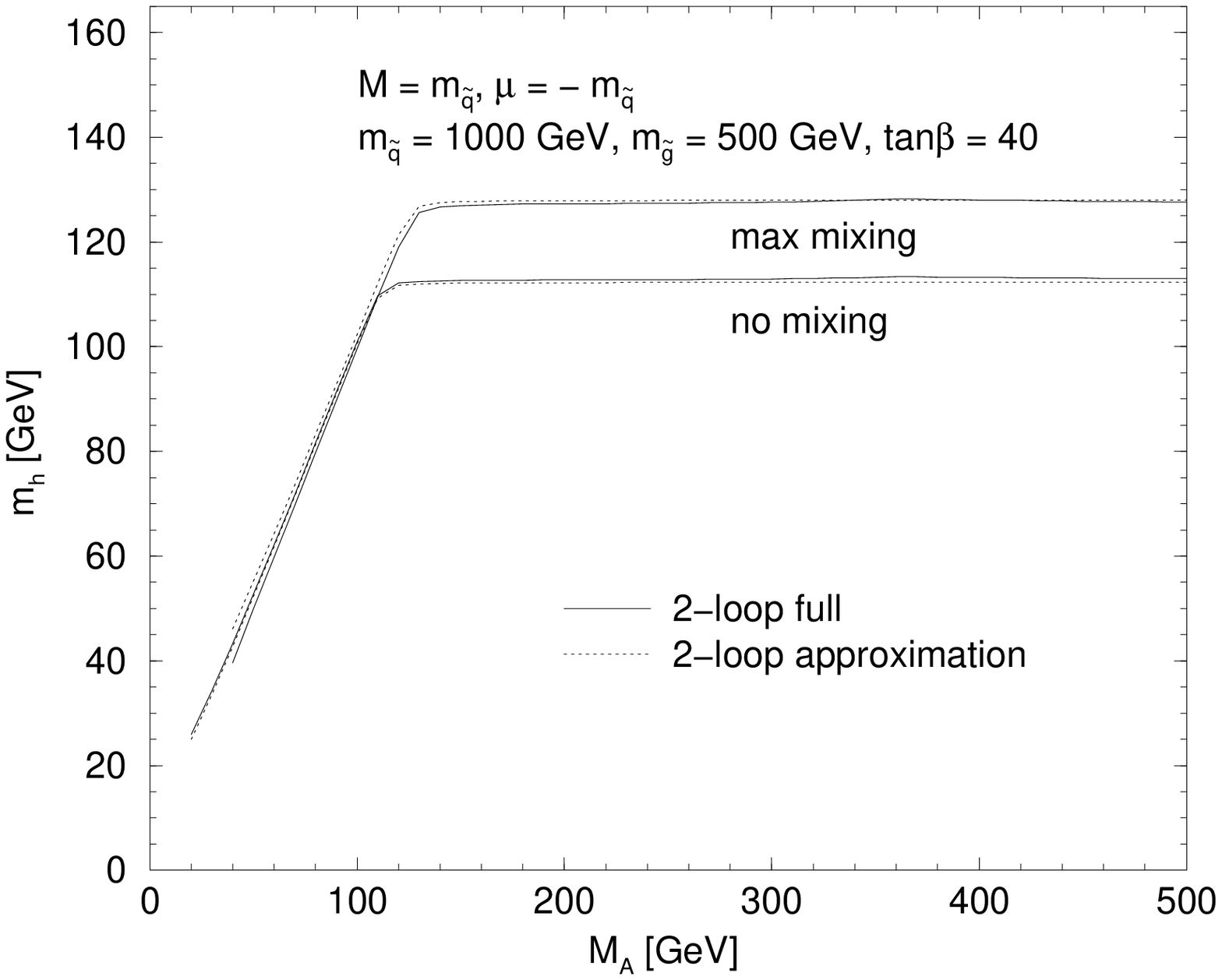,width=7cm,height=5.4cm}}
\end{center}
\vspace{-1.5em}
\caption[]{
$\mh$ as a function of $\MA$, calculated from the full formula and
from the approximation formula for $\msq = 1000 \gev, 
\mgl = 500 \gev$ and $\tb = 1.6$ or $40$.
}
\label{fig:mh_MA}
\end{figure}
%
\begin{figure}[ht!]
\begin{center}
\mbox{
\psfig{figure=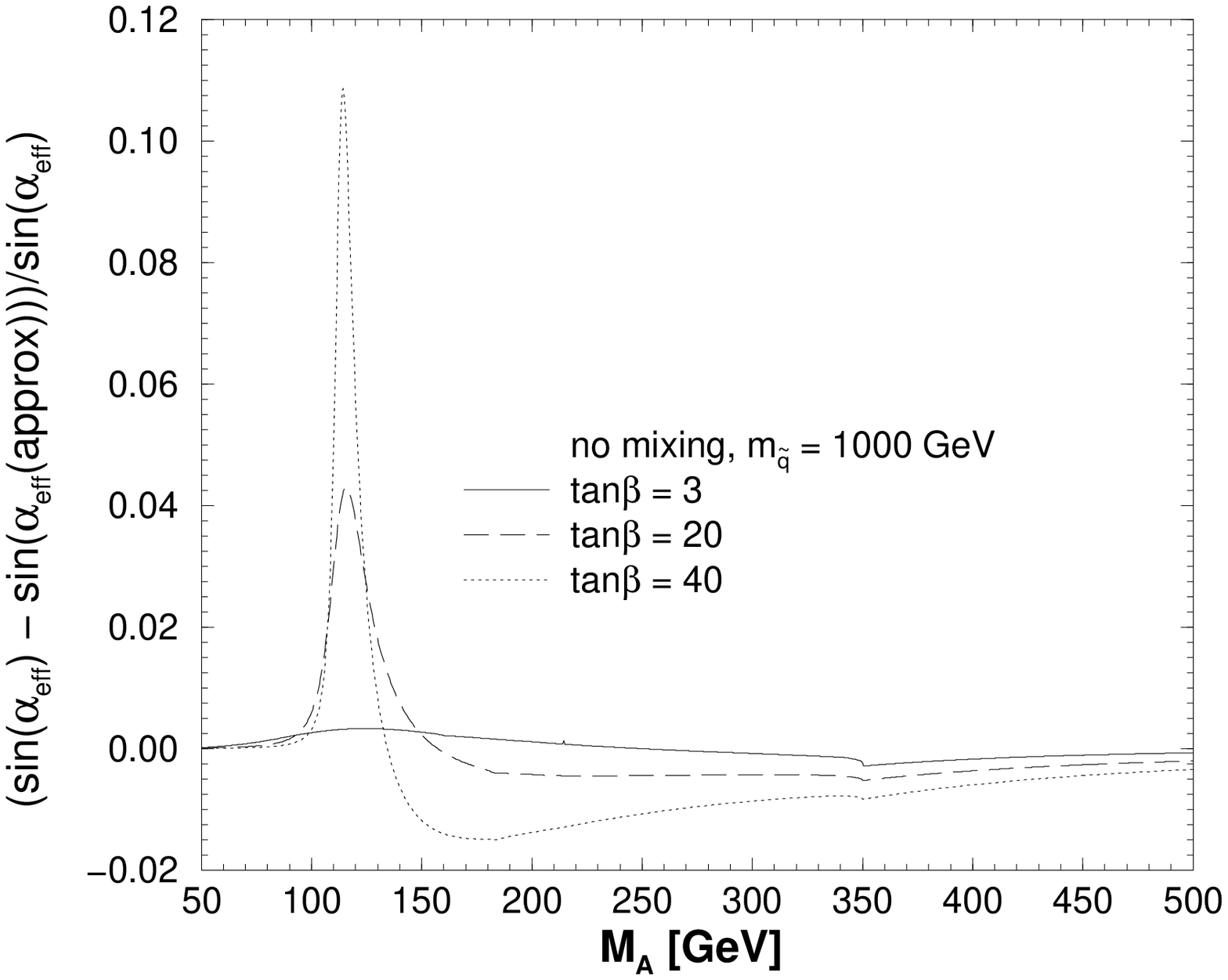,width=7cm,height=5.4cm}}
\hspace{1em}
\mbox{
\psfig{figure=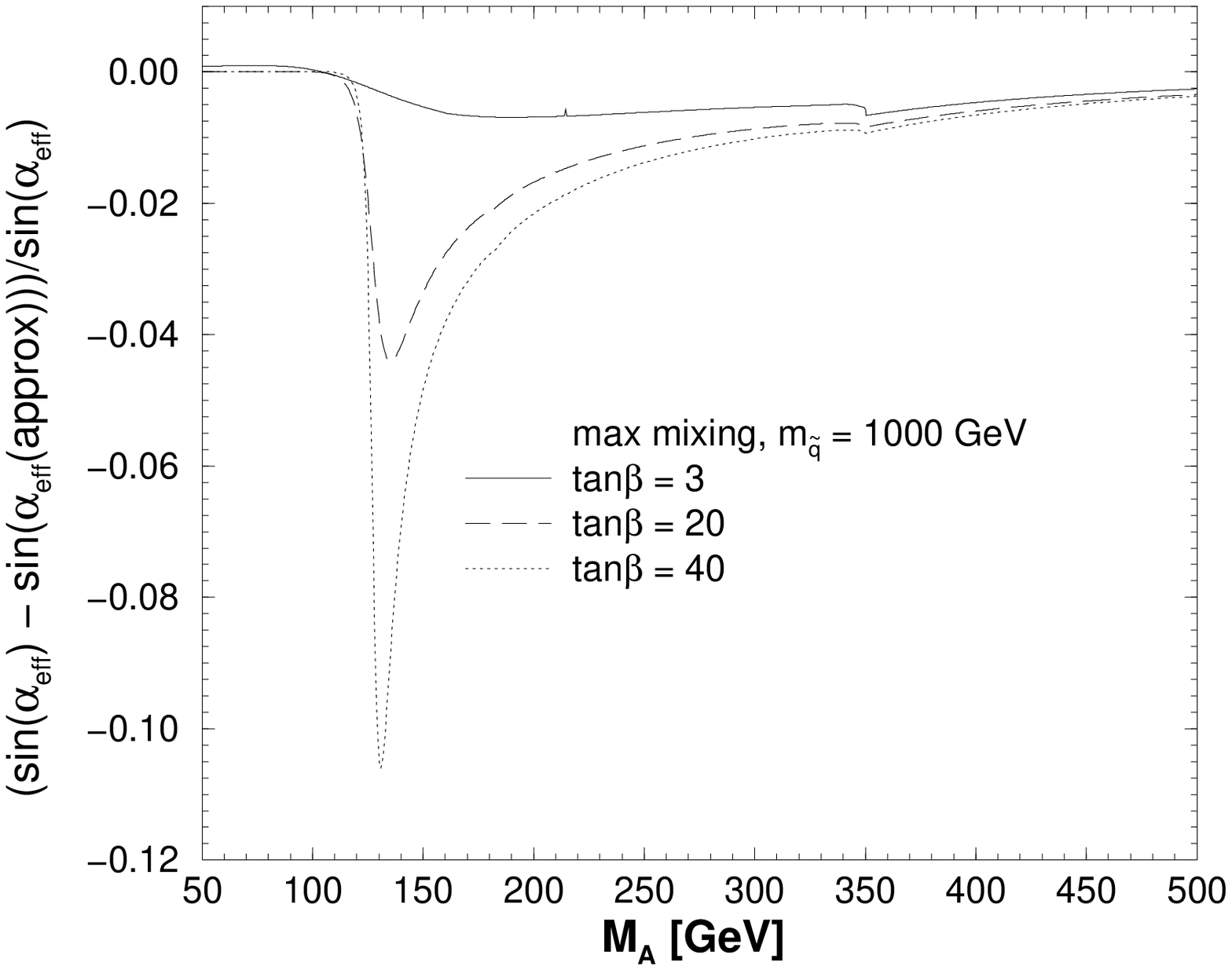,width=7cm,height=5.4cm}}
\end{center}
\vspace{-1.5em}
\caption[]{
The relative difference $(\sin\aeff - \sin\aeffapprox)/\sin\aeff$  
is shown as a function
of $\MA$ for three values of $\tb$ in the no mixing and the maximal
mixing scenario. The other parameters are $\mu = -100 \gev$, 
$M = \msq$, $\mgl = 500 \gev$ and $A_b = A_t$.
}
\label{fig:saeff_MA}
\vspace{-2em}
\end{figure}

For the Standard Model parameters we use 
$\MZ = 91.187 \gev,~\MW = 80.39 \gev,~
G_F = 1.16639\times 10^{-5} \gev^{-2},~\als(\mt) = 0.1095$, and
$\mt = 175 \gev$. In the numerical evaluation we
have furthermore chosen the trilinear couplings in the scalar top and
bottom sector to be $A_b = A_t$. This fixes together with the
choice of $\mu$ (the Higgs mixing parameter) the mixing in the scalar
bottom sector. 
The parameter $M$ appearing in the plots is the $SU(2)$ gaugino mass
parameter, it enters as an independent parameter in the full result
only, see \citere{mhiggslong}. 

In \citeres{mhiggslong,mhiggslle} it has been shown that the lightest
Higgs boson mass $\mh$ as a function of $\Xt$ reaches a maximum at
around $|\Xt/\msq| \approx 2$. This case we refer to as 'maximal
mixing'. A minimum of $\mh$ is reached for $\Xt \approx 0$, which we
refer to as 'no mixing'.

\smallskip
\reffi{fig:mh_mq} displays the dependence of $\mh$ on $\msq$ for the cases
of no mixing and maximal mixing, and we have set $\MA = 500 \gev$.
For $\tb$ we have chosen two typical values: $\tb = 1.6$ as a low and
$\tb = 40$ as a typical high value.
Very good agreement is found  in the
no-mixing scenario as well as in the maximal-mixing scenario,
the deviations lie below $2 \gev$.

The dependence on $\MA$ is shown in \reffi{fig:mh_MA}. 
The quality of the approximation is typically better than $1 \gev$ for
the no-mixing case and better than $2 \gev$ for the maximal-mixing
case. Only for very small (and experimentally already excluded) values
of $\MA$ a deviation of $5 \gev$ occurs. 
The peaks in the plot for $\tb = 1.6$ in the full result are due to
the threshold  $\MA = 2\,\mt$ in the
\onel\ contribution, originating from the top-loop diagram in
the $A$~self-energy. This peak does not occur in the approximation
formula (where the momentum dependence of the $A$ self-energy
has been neglected) and can thus lead to a larger deviation around the
threshold. 

In \reffi{fig:saeff_MA} we display the quality of the short analytical
approximation formula for the effective mixing angle. The relative
difference 
between $\aeff$ obtained from the full calculation (i.e.\ calculating
the renormalized Higgs self-energies $\hSi_s(0)$ without any
approximation) and the angle obtained with the help of the
approximated Higgs self-energies, denoted as $\aeffapprox$ is shown as
a function of $\MA$. 
We use three values of $\tb$, $\tb = 3, 20, 40$.
Sizable deviations
occur only in the region $100 \gev \le \MA \le 150 \gev$, especially
for large $\tb$. In this region of parameter space the values of $\mh$
and $\mH$ are very close to each other. This results in a high
sensitivity to small deviations in the Higgs boson self-energies
entering the Higgs-boson mass
matrix~(\ref{higgsmassmatrix}),~(\ref{alphaeff}). Otherwise the
relative difference stays below~3\%.


\section{The Fortran program \fhf}


The complete program \fhf\ consists of about 1300 lines Fortran code.
The executable file fills about 65 KB disk space.
The calculation for one set of parameters, including the $\De\rho$
constraint, takes about $2 \times 10^{-5}$ seconds on 
a Sigma station (Alpha processor, 600 MHz processing speed, 512 MB RAM).
The program can be obtained from the \fh\ home page:\\
{\tt http://www-itp.physik.uni-karlsruhe.de/feynhiggs}~.\\
Here the code is available,
together with a short instruction, information about bug fixes etc.

\fhf\ consists of a front-end ({\tt program FeynHiggsFast}) and the
main part where the evaluation is performed (starting with 
{\tt subroutine feynhiggsfastsub}). The front-end can be manipulated
by the user at will, whereas the main part should not be changed. In
this way \fhf\ can be accommodated as a subroutine to existing
programs%
\footnote{
This has been carried out, for example, for the program
HDECAY~\cite{hdecay,spirix}, into which \fhf\ has been incorporated as
a subroutine.
}
, thus providing an extremely fast and for many purposes sufficiently
accurate evaluation for the masses and
mixing angles in the MSSM Higgs sector. 

The input of \fhf\ are the low energy SUSY parameters, listed in
\refta{tab:mssmparameters}. Concerning the $\Stop$-sector, the user
has the option to enter either the physical parameters, i.e.\ the
masses and the mixing angle ($\mste$, $\mstz$ and $\sintt$) or the
unphysical parameters $\MstL$, $\MstR$ and $\Xt$. From these input
parameters \fhf\ calculates the masses and the mixing angle of the
neutral $\cp$-even Higgs sector, as well as the mass of the charged
Higgs boson and the \rp.

\begin{table}[ht!]
\renewcommand{\arraystretch}{1.5}
\begin{center}
\begin{tabular}{|c||c||c|} \hline
{\bf input for \fhf} & {\bf MSSM expr.} & 
{\bf internal expr. in \fhf}
\\ \hline \hline
{\tt tan(beta)}          & $\Tb$                     & {\tt ttb} \\
{\tt Msusy\_top\_L  }    & $\MstL$                   & {\tt msusytl} \\
{\tt Msusy\_top\_R}      & $\MstR$                   & {\tt msusytr} \\
{\tt MtLR }              & $\Xt \equiv \Mtlr$        & {\tt mtlr} \\
{\tt MSt2}               & $\mstz$                   & {\tt mst2} \\
{\tt delmst}             & $\delmst = \mstz - \mste$ & {\tt delmst} \\
{\tt sin(theta\_stop)}   & $\sintt$                  & {\tt stt} \\
{\tt MT}                 & $\mt$                     & {\tt mmt} \\
{\tt Mue}                & $\mu$                     & {\tt mmue} \\
{\tt MA}                 & $\MA$                     & {\tt mma} \\
\hline
\end{tabular}
\renewcommand{\arraystretch}{1}
\caption[]{The meaning of the different MSSM variables which have to
be entered into \fhf.}
\label{tab:mssmparameters}
\end{center}
\end{table}


\section{Conclusions}

\fhf\ is a Fortran code for the calculation of masses and mixing
angles in the Higgs sector of the MSSM. In addition it evaluates the
SUSY corrections to the \rp, arising from the scalar top and bottom sector.
Concerning the evaluation in the neutral $\cp$-even Higgs sector, 
\fhf\ is based on a simple analytic formula, derived from a more
complete result obtained in 
the Feynman-diagrammatic approach. 
Beyond the \onel\ level, \fhf\ contains the dominant corrections in
$\oaas$ and further subdominant contributions.
The accuracy of the approximation compared to the full result is better
than $2 \gev$ for most parts of the MSSM parameter space.

The program is available via the WWW page\\
{\tt http://www-itp.physik.uni-karlsruhe.de/feynhiggs}.

The code consists of a front-end and a subroutine. The front-end can
be manipulated at the user's will. By accommodating the front-end,
\fhf\ can serve as a subroutine to existing programs, thus providing
an extremely fast and reasonably accurate evaluation of the masses and 
mixing angles in the MSSM Higgs sector. These 
can then be used as inputs for further computations.


\bigskip
\subsection*{Acknowledgements}
S.H. thanks the organizers of the Les Houches workshop for the
inspiring and relaxed atmosphere.




\end{document}